\documentstyle[pra,aps,epsfig]{revtex}
\begin{document}
\flushbottom
\draft
\title{Higher-order mutual coherence of optical and matter waves}
\author{G.A. Prataviera$^1$, J. Zapata$^{1,2}$ , and P. Meystre$^1$}
\address{$^1$Optical Sciences Center, University of Arizona, Tucson, Arizona
85721\\
$^2$Departamento de F\'\i sica, Universidad Nacional de Colombia, Bogot\'a.}
\maketitle
\begin{abstract}
We use an operational approach to discuss ways to measure the higher-order
cross-correlations between optical and matter-wave fields. We pay particular
attention to the fact that atomic fields actually consist of composite particles
that can easily be separated into their basic constituents by a detection
process such as photoionization. In the case of bosonic fields, that we
specifically consider here, this leads to the appearance in the
detection signal of exchange contributions due to both the composite bosonic
field and its individual fermionic constituents.  We also show how time-gated
counting schemes allow to isolate specific contributions to the signal,
in particular involving different orderings of the Schr\"odinger and
Maxwell fields.
\end{abstract}
\section{Introduction}
The experimental realization of Bose-Einstein condensation in dilute atomic
systems \cite{r1,r2,r3,r4,r5} has opened up exciting new directions of research
in atomic, molecular and optical physics. Recent developments of special
interest in the context of the present paper include the first experimental
demonstration of the four-wave mixing of de Broglie waves in a Sodium
condensate \cite{r6}, the generation of dark solitons by optically imposing a
phase shift on a condensate wave function \cite{ert99,phi99}, the realization
of several types of atom lasers \cite{mew97,and98,r7,r8},
and, most importantly perhaps, the demonstration of superradiant
Rayleigh scattering in a condensate \cite{r9}. This
latter experiment is particularly significant in that it is the first example
of a four-wave mixing process involving dynamically two optical waves and
two atomic matter waves.

In recent work \cite{r10}, we initiated a study of the mutual coherence between
optical and matter waves, discussing possible measurement schemes, and illustrating
our results in a simple example. The goal of the present paper is to extend
these results to the study of higher-order coherence functions.
This study is motivated by several reasons: First, it is well
established in optics that higher-order detection schemes are required to
test certain aspects of the nonlocal predictions of quantum mechanics.
\cite{r11} It would be useful to carry out similar experiments with massive
particles. Second, the generation of a quantum entanglement between optical
and matter waves is of much interest in applications such as the optical
manipulation and control of Schr\"odinger fields. In particular, a number
of sophisticated techniques have been developed to generate optical waves of
prescribed statistical properties. It would be desirable to develop
methods to transfer them to matter waves, while possibly amplifying
them in the process. This might for example open up the way to the
generation of atomic Fock states of known atomic number, which are of great
potential interest in matter-wave interferometry. \cite{BouKas97} More generally,
the interplay between Maxwell and Schr\"odinger waves is the cornerstone
of most potential applications of atom optics, from situations such as
lenses, gratings and mirrors where light plays a passive role to situations
of dynamical coupling, including matter-wave amplifiers \cite{ket99z}
and ultra-sensitive atom-optical detectors. These developments require one
to characterize not just the statistical properties of the optical and
matter fields individually, but also their cross-coherence properties.

The general detection scheme that we consider consists of one or more standard
photodetectors for the optical field, together with matter-wave detectors
which operate by annihilating an atom via photoionization. As a result,
electrons and ions are produced, and the resulting electron current is measured.
The atomic detection, while on the surface similar to photodection, is therefore
fundamentally different in that the atoms, which  are composite particles,
are not so much destroyed by the detection process as transformed into a pair
of two new particles. For example, if the atoms under consideration are
composite bosons before detection, the resulting ions are fermions. The quantum
statistics of the resulting particles is important in those situations
involving exchange terms, as we discuss in detail later on.

The paper is organized as follows: Section II reviews the model of joint
atomic and optical detection of Ref. \cite{r10} and establishes the notation.
Section III, which is the central part of the paper, discusses the kinds
of correlation functions that can be detected. It accounts in detail for
the quantum statistics of the ions and electrons resulting from photoionization,
including the effects of particle exchange, and illustrates how different 
orderings can be achieved by a proper time-gating
of the detectors. Section IV illustrates these results in a simple atom
optics example, and Section V is a summary and conclusion.

\section{Detection scheme}

We consider the joint detection of the coherence
properties of dynamically coupled optical and matter-wave fields, the
latter one being assumed to be bosonic for concreteness. In Ref. \cite{r10},
we proposed a detection scheme that can achieve this goal. It consists of
a series of standard photodetectors, which we call in the following
{\em Maxwell detectors} and of matter-wave {\em Schr\"odinger
detectors.} The Schr\"odinger detectors operate by tightly focussing
a laser beam on the atoms, ionizing them, and measuring the resulting
electron current. We describe the optical field by its electric field
operator ${\bf E}({\bf r},t)$ and the matter-wave field by the
(multicomponent) Schr\"odinger field operator ${\bf \Psi}({\bf r},t)$, which
satisfies the bosonic commutation relation
\begin{equation}
[{\bf \Psi}_{\alpha}({\bf r} ),{\bf \Psi}_{\beta}^{\dag}({\bf r')}]=
\delta_{\alpha\beta}\delta(\bf{r}-\bf{r'}).
\label{e1}
\end{equation}
Here ${\bf r}$ is the center-of-mass coordinate and the indices
$\alpha$ and $\beta$ label the internal (electronic) state of the atoms.

The dynamics of the compound system is governed by the Hamiltonian
\begin{equation}
{\cal H}_0={\cal H}_m+{\cal H}_s+{\cal V}
\label{e2}
\end{equation}
where ${\cal H}_m$ and ${\cal H}_s$ describe the evolution of the light
field and the matter-wave field, respectively, while $\cal V$ is the
interaction between these two fields, typically the electric dipole interaction.

In Ref. \cite{r10}, we considered the measurement of the lowest-order
joint-correlations between the two fields. In that case, just one
photodetector and one atom detector were required. In contrast, the
measurement of higher-order correlations involves the use of several light
and matter-wave detectors.

As in Glauber's photodetection theory \cite{glauber}, the photodetectors are modeled as
single two-level atoms whose excited state is in the continuum, corresponding
to photoionization. The interaction of the system with $N$ such detectors
is described as usual by the electric dipole interaction Hamiltonian
\begin{equation}
\label{e3}
{\cal V}_m=\sum_{j=1}^{N}{\bf d}_j\cdot {\bf E}({\bf r}_j)
\end{equation}
where ${\bf d}_j$ is the electric dipole moment between the ground electronic
state and the continuum state of the $j^{th}$ detector, located
at the position ${\bf r}_j$. We assume for simplicity that the dipole
moments are real and parallel to the field. Similarly, the interaction
between the matter-wave field and the $M$ Schr\"odinger detectors,
assumed to be point-like and at locations ${\bf r}_j$, is given by
\begin{equation}
{\cal V}_s =\hbar\sum_{j=N+1}^{N+M}\sum_{\nu } [\Omega_{\nu } ({\bf r}_{j})
{\bf \Psi}_{\nu}^{\dag}({\bf r}_j )
{\bf \Psi}({\bf r}_j )e^{-i\omega_Lt}+ h.c.].
\label{e4}
\end{equation}
Here, ${\bf \Psi}$ and ${\bf \Psi}_{\nu}$ are the field operators associated
with atoms in the (ground state) sample \footnote{We consider only scalar
fields in the following for notational simplicity. The extension to the
detection of multicomponent Schr\"odinger fields is straightforward.}
and the atomic continuum state $|\nu \rangle$, respectively,
$\Omega_{\nu }$ is the coupling constant proportional to the
Rabi frequency of the ionizing lasers of frequency
$\omega_L$. As pointed out in Ref. \cite{elena}, the assumption of
point-like detectors is justified for atomic samples at temperatures well
below the recoil limit, in which case the ionizing lasers
can easily be focused to a spot size  much smaller than the thermal
de Broglie wavelength.

The atoms in the Schr\"odinger field ${\bf \Psi}({\bf r})$ are actually composite
bosons which are photodissociated by the laser into an electron and an ion.
Hence, the creation operator ${\bf \Psi}_\nu^\dag({\bf r})$ should actually
be understood as the product of two fermionic creation operators. Since what
is detected is the electron current, it is important to treat this
property of the system properly. We proceed by expanding the continuum
field ${\bf \Psi}_\nu({\bf r})$ in terms of plane waves of momentum ${\bf q}$
as
\begin{equation}
\label{e6}
{\bf \Psi}_{\nu}({\bf r})=
\int d^3q \phi_{{\bf q}} ({\bf r}) b_{\nu {\bf q}}
\end{equation}
where
\begin{equation}
\phi_{{\bf q}}({\bf r}) = \frac{1}{(2\pi )^{3/2}}e^{i{\bf q}\cdot {\bf r}} ,
\end{equation}
\begin{equation}
\omega_{\nu q}= \hbar {\bf q}^{2}/2M +\omega_{\nu},
\end{equation}
and
\begin{equation}
[b_{\nu {\bf q}},b_{\nu ' {\bf q}'}^{\dagger}]=
\delta ({{\bf q}-{\bf q}'})\delta_{\nu \nu '},
\end{equation}
the corresponding Hamiltonian being
\begin{equation}
\label{e5}
{\cal H}_I = \hbar\sum_{\nu} \int d^3q \omega_{\nu q} b_{\nu {\bf q}}^\dagger
b_{\nu {\bf q}} .
\end{equation}
We now decompose the atomic creation operators $b^\dag_{\nu {\bf q}}$ in
terms of products of electron and ion creation operators. We assume for
concreteness that the atoms under consideration are spin-zero particles, so that
the spins of the resulting electrons and ions are opposite. As a result of the
large difference between the ion and the electron masses, the kinetic energy of the
system is carried almost entirely by the ion. If furthermore the photon energy is
insufficient to excite the ion above its ground electronic state, we have that
\begin{equation}
b^\dag_{\nu {\bf q}} = \sum_s a^\dag_{{\bf q},s}
\int d^3 k \varphi^{*}_\nu({\bf k}) c^\dag_{\nu, -s}({\bf k})
\label{b}
\end{equation}
where $\varphi_\nu({\bf k})$ are momentum-space electron wave functions.
The sum appearing in this equation is over the spin states $s = \pm 1/2$, and
the ion and electron annihilation operators, $a_{{\bf q},s}$ and
$c_{\nu,s}({\bf k})$ respectively, satisfy the Fermi anticommutation relations
\begin{equation}
[a_{{\bf q},s},a^\dag_{{\bf q}',s'}]_+ = \delta_{s,s'} \delta
({\bf q} - {\bf q'})
\end{equation}
and
\begin{equation}
[c_{\nu,s}({\bf k}),c^\dag_{\nu',s'}({\bf k'})]_+
= \delta_{s,s'} \delta_{\nu, \nu'} \delta ({\bf k}-{\bf k}').
\end{equation}
Combining Eqs. (\ref{b}) and (\ref{e6}) gives
\begin{equation}
\label{ie}
{\bf \Psi}_{\nu}({\bf r}) =\sum_s \int d^3q \phi_{{\bf q}} ({\bf r})
a_{{\bf q},s} \int d^3 k \varphi_\nu({\bf k}) c_{\nu, -s}({\bf k}) .
\label{expand}
\end{equation}
The Schr\"odinger detectors measure the electron current, but are insensitive
to the final state of the ions, which will therefore be summed over.

The total Hamiltonian describing the coupling of the system to the
detectors is finally
\begin{equation}
\label{e7}
{\cal V}_d= {\cal V}_m +{\cal V}_s
\end{equation}
and the full Hamiltonian including system, detectors and interaction
is
\begin{equation}
\label{e8}
{\cal H}={\cal H}_0+{\cal H}_{I}+{\cal H}_{A}+{\cal V}_d
\end{equation}
where ${\cal H}_{A}$ is the Hamiltonian of the Maxwell detectors,
modeled as usual as two level-atoms.

\section{Counting signal}
The probability amplitude of exciting all detectors
is given to lowest order by $(N+M)$-th order perturbation theory. To
that order, the probability amplitude for the transition from the initial
state $|i\rangle$ to a final state $|f\rangle$ is given by
\begin{equation}
\label{b1}
\left ({-i\over{\hbar}}\right)^{N+M}
\int_{0}^{t}d\tau_1\int_{0}^{t}d\tau_2...
\int_{0}^{t}d\tau_{N+M}
\langle f|{\cal V}_d(\tau_1){\cal V}_d(\tau_2)...
{\cal V}_d(\tau_{\tiny{N+M}})|i\rangle
\end{equation}
where $\tau_1 >\tau_2 >...>\tau_{N+M}$, and the interaction with the
detectors is now expressed in the interaction picture as
\begin{equation}
\label{b2}
{\cal V}_d (\tau )\to e^{i({\cal H}_{0}+{\cal H}_{d})\tau }
{\cal V}_d (\tau )
e^{-i({\cal H}_{0}+{\cal H}_{d})\tau },
\end{equation}
where ${\cal H}_{d}={\cal H}_{I}+{\cal H}_{A}$ is the free Hamiltonian of the
detectors.

When substituted into Eq. (\ref{b1}), the interaction Hamiltonian (\ref{e7})
yields $(N+M)^{(N+M)}$ terms. It is well known that for photodetectors
operating by absorption, only the positive frequency part of the
electric field operator contributes to the signal. A similar situation also
occurs for the Schr\"odinger detectors provided that no ion 
or electron is initially present,
i.e., the field ${\bf {\Psi}}_\nu({\bf r})$ is initially in the vacuum state.
If that is the case, the only terms contributing to the detection signal are
those proportional to ${\bf \Psi}_{\nu}^{\dag}({\bf r}_{j} )
{\bf \Psi}({\bf r}_j )e^{-i\omega_Lt}$, which result in the detection
of normally ordered correlation functions of the Schr\"odinger field.
Further neglecting terms involving multiple ionization at a single detector
reduces the probability amplitude
(\ref{b1}) to
\begin{equation}
\label{b3}
\left ({-i\over{\hbar}}\right)^{N+M}
\int_{0}^{t}d\tau_1\int_{0}^{t}d\tau_2...\int_{0}^{t}d\tau_{\tiny{N+M}}
\sum_{{\it all\,\,permutations}}
\langle f|{\cal V}_1^{(+)}(\tau_1){\cal V}_2^{(+)}(\tau_2)
...{\cal V}_{\tiny{N+M}}^{(+)}(\tau_{\tiny{N+M}})|i\rangle .
\end{equation}
In this expression, ${\cal V}_i$ can correspond to either the detection of
a photon, or to the ionization of an atom. All permutations of such
detection events such that a total of $N$ photodetections of the optical
field and $M$ photoionizations of the Schr\"odinger field take place must be
included in the sum. The ``+'' superscrit refers to the familiar
``positive frequency part'' in the Maxwell detectors and to the ordering of
the Schr\"odinger detectors discussed before Eq. (\ref{b3}). Hence this
expression includes $(N+M)!$ terms. Since the optical
and atomic field operators do not generally commute at different times,
it is not possible to regroup them into just one contribution. However, it is
possible to use a time-gated detection scheme \cite{cohen} to separately
detect the various correlation functions of Eq. (\ref{b3}), as we now
discuss.

We consider the experimental situation where the $i$-th detector is
turned on at $t=0$ and turned off at some later time $t_i$, so that
${\cal V}_d$ becomes
\begin{equation}
\label{b4}
{\cal V}_d(t)= \sum_{i=1}^{N+M}{\cal V}_i
\left [ \Theta(t) - \Theta(t-t_i)\right ].
\end{equation}
where $\Theta(t)$ is the Heaviside function. A given ordering of
the switch-off times, for example $t_1<t_2...<t_{N+M}<t$ results in just one
term in Eq. (\ref{b3}) contributing to the counting rate.\cite{cohen}

As is the case in standard optical photodetection theory \cite{glauber},
we can safely assume that the photocurrents generated in the Maxwell detectors
are independent and distinguishable. However, more care must be taken when
dealing with the Schr\"odinger detectors: Because the de Broglie wavelength
of the ultracold atoms can be very large, it is not correct in general to
assume that the electrons generated by the photoionization process are
distinguishable. Hence, it is necessary to account for the
various exchange terms arising in the detection signal \cite{elena}.

We illustrate how this work in the case of a detection scheme involving two
Maxwell detectors at positions ${\bf r}_1$ and ${\bf r}_2$ and turned off at
times $t_1$ and $t_2$, respectively, as well as two Schr\"odinger
detectors at ${\bf r}_3$ and ${\bf r}_4$ and closed at times $t_3$ and $t_4$.
For the time ordering $t_1<t_2<t_3<t_4<t$, the only term resulting from the
transition probability amplitude (\ref{b3}) that contributes to the
counting rate turns out to be
\begin{eqnarray}
\label{b5}
{\cal P}(t)  &\approx & {1\over\hbar^8}\sum_{\nu_1\nu_2\nu_3\nu_4}
\Omega_{\nu_1}^*({\bf r}_3)\Omega_{\nu_2}^*({\bf r}_4)
\Omega_{\nu_3}({\bf r}_4)\Omega_{\nu_4}({\bf r}_3) \nonumber \\
&\times& \int_{0}^{t_4}d\tau_4\int_{0}^{t_4}d\tau'_4...
\int_{0}^{t_1}d\tau_1\int_{0}^{t_1}d\tau_1'
e^{-i\omega_{L} (\tau_4-\tau_4'+\tau_3-\tau_3')}\nonumber \\
&\times & \langle d_1(\tau_1')d_1(\tau_1)d_2(\tau_2')d_2(\tau_2)\rangle
\langle {\bf \Psi}_{\nu_1}({\bf r}_3,\tau_3')
{\bf \Psi}_{\nu_2}({\bf r}_4,\tau'_4) {\bf \Psi}_{\nu_3}^{\dagger}
({\bf r}_4,\tau_4){\bf \Psi}_{\nu_4}^{\dagger}
({\bf r}_3,\tau_3)\rangle \nonumber \\
 &\times & \langle {\bf E}^{(-)}({\bf r}_1,\tau'_1)
{\bf E}^{(-)}({\bf r}_2,\tau'_2)
{\bf \Psi}^{\dagger}({\bf r}_3,\tau'_3){\bf \Psi}^{\dagger}({\bf r}_4,\tau'_4)
{\bf \Psi}({\bf r}_4,\tau_4){\bf \Psi}({\bf r}_3,\tau_3)
{\bf E}^{(+)}({\bf r}_2,\tau_2)){\bf E}^{(+)}({\bf r}_1,\tau_1)\rangle
\end{eqnarray}
where we have introduced as usual a sum over a complete set of final states.
Note that in deriving this expression, we have factorized the expectation values
into products of detector operators and of system operators correlation 
functions. This is appropriate in the absence of quantum entanglement 
between the states of the detectors and of the system to be characterized.

In order to properly account for the quantum statistics of the detector
correlation function
\begin{equation}
C^{(2)}_d \equiv \langle {\bf \Psi}_{\nu_1}({\bf r}_3,\tau_3')
{\bf \Psi}_{\nu_2}({\bf r}_4,\tau'_4) {\bf \Psi}_{\nu_3}^{\dagger}
({\bf r}_4,\tau_4){\bf \Psi}_{\nu_4}^{\dagger}
({\bf r}_3,\tau_3)\rangle ,
\end{equation}
we reexpress it using Eq. (\ref{ie}) as
\begin{eqnarray}
C^{(2)}_d &=&
\sum\!\!\!\!\!\!\!\!\int
\exp\left \{ -i\left [(\omega_{{\bf q}_1}+\omega_{\nu_1})\tau_3'+ (\omega_{{\bf q}_2}
+\omega_{\nu_2})\tau_4'-(\omega_{{\bf q}_3}+\omega_{\nu_3}) \tau_4
-(\omega_{{\bf q}_4}+ \omega_{\nu_4}) \tau_3\right ] \right \} \nonumber \\
&\times & \phi_{{\bf q}_1}({\bf r}_3)
\phi_{{\bf q}_2}({\bf r}_4) \phi_{{\bf q}_3}^{*}({\bf r}_4)
\phi_{{\bf q}_4}^{*}({\bf r}_3)\varphi_{\nu_1}({\bf k}_1)
\varphi_{\nu_2}({\bf k}_2)\varphi_{\nu_3}^{*}({\bf k}_3)
\varphi_{\nu_4}^{*}({\bf k}_4)\nonumber \\
&\times & \langle a_{{\bf q}_1,s_1} a_{{\bf q}_2,s_2}a^{\dagger}_{{\bf q}_3,s_3}
a^{\dagger}_{{\bf q}_4,s_4}\rangle \langle c_{\nu_1,-s_1}({\bf k}_1)
c_{\nu_2,-s_2}({\bf k}_2)c^{\dagger}_{\nu_3,-s_3}({\bf k}_3)
c^{\dagger}_{{\nu_4,-s_4}}({\bf k}_4)\rangle
\label{b6}
\end{eqnarray}
where the sum is over a set of discrete and continuous indices
$s, {\bf q}$ and ${\bf k}$ and we have assumed that the electron and ion
dynamics are governed by their free Hamiltonians. Since the ion and electron
fields are in the vacuum state, the expectation value of the ions and electrons
operators can be easily evaluated using their fermion anticommutation
relations to give
\begin{equation}
\label{commutator1}
\langle a_{{\bf q}_1,s_1}a_{{\bf q}_2,s_2}a^{\dagger}_{{\bf q}_3,s_3}
a^{\dagger}_{{\bf q}_4,s_4}\rangle =\delta({\bf q}_1-{\bf q}_4)
\delta({\bf q}_2-{\bf q}_3)\delta_{s_1,s_4}\delta_{s_2,s_3}
-\delta({\bf q}_1-{\bf q}_3)
\delta({\bf q}_2-{\bf q}_4)\delta_{s_1,s_3}\delta_{s_2,s_4}
\end{equation}
and
\begin{eqnarray}
\label{commutator2}
\langle c_{\nu_1,-s_1}({\bf k}_1)
& &c_{\nu_2,-s_2}({\bf k}_2)c^{\dagger}_{\nu_3,-s_3}({\bf k}_3)
c^{\dagger}_{{\nu_4,-s_4}}({\bf k}_4)\rangle = \nonumber \\
& &\delta({\bf k}_1-{\bf k}_4)
\delta({\bf k}_2-{\bf k}_3)\delta_{\nu_1,\nu_4}\delta_{\nu_2,\nu_3}
\delta_{s_1,s_4}\delta_{s_2,s_3}
-\delta ({\bf k}_1-{\bf k}_3)
\delta ({\bf k}_2 -{\bf k}_4)\delta_{\nu_1,\nu_3}\delta_{\nu_2,\nu_4}
\delta_{s_1,s_3}\delta_{s_2,s_4}.
\end{eqnarray}
As a result, $C^{(2)}_d$ contains four terms, associated with direct and 
exchange contributions to the signal at the Schr\"odinger detectors.

In case both optical and matter-wave detectors are broadband \cite{cohen},
with response spectra centered at in $\omega_m$ and $\omega_s$ respectively,
Eq. (\ref{b5}) reduces to
\begin{eqnarray}
{\cal P}(t) &\approx &\eta_m^2 \left \{
\eta_s({\bf r}_3)\eta_s({\bf r}_4) \int_{0}^{t_4}d\tau_4
 \int_{0}^{t_3}d\tau_3\int_{0}^{t_2}d\tau_2 \int_{0}^{t_1}d\tau_1 \right.
 \nonumber \\
&\times& \langle {\bf E}^{(-)}({\bf r}_1,\tau_1)
{\bf E}^{(-)}({\bf r}_2,\tau_2)
{\bf \Psi}^{\dagger}({\bf r}_3,\tau_3){\bf \Psi}^{\dagger}({\bf r}_4,\tau_4)
{\bf \Psi}({\bf r}_4,\tau_4){\bf \Psi}({\bf r}_3,\tau_3)
{\bf E}^{(+)}({\bf r}_2,\tau_2)){\bf E}^{(+)} ({\bf r}_1,\tau_1)\rangle
\nonumber \\
&+& \eta_s({\bf r}_3,{\bf r}_4)\eta_s({\bf r}_4 ,{\bf r}_3)
\int_{0}^{t_4}d\tau_4  \int_{0}^{t_3}d\tau_3 \int_{0}^{t_2}
d\tau_2 \int_{0}^{t_1} d\tau_1
\nonumber \\
&\times& \langle {\bf E}^{(-)}({\bf r}_1,\tau_1)
{\bf E}^{(-)}({\bf r}_2,\tau_2 )
{\bf \Psi}^{\dagger}({\bf r}_3,\tau_3 ){\bf \Psi}^{\dagger}({\bf r}_4,\tau_4 )
{\bf \Psi}({\bf r}_4,\tau_3){\bf \Psi}({\bf r}_3,\tau_4)
{\bf E}^{(+)}({\bf r}_2,\tau_2)){\bf E}^{(+)}({\bf r}_1,\tau_1)\rangle
 \nonumber \\
&- & \eta_x({\bf r}_3,{\bf r}_4)\int_{0}^{t_3}d\tau_3
\int_{0}^{t_2}d\tau_2 \int_{0}^{t_1}d\tau_1
\nonumber \\
&\times&  \left. \langle {\bf E}^{(-)}({\bf r}_1,\tau_1)
{\bf E}^{(-)}({\bf r}_2,\tau_2 )
{\bf \Psi}^{\dagger}({\bf r}_3,\tau_3){\bf \Psi}^{\dagger}({\bf r}_4,\tau_3 )
{\bf \Psi}({\bf r}_4,\tau_3){\bf \Psi}({\bf r}_3,\tau_3 )
{\bf E}^{(+)}({\bf r}_2,\tau_2 ){\bf E}^{(+)}({\bf r}_1,\tau_1 )\rangle \right \}
\label{b7}
\end{eqnarray}
where the efficiency of the photodetectors
\begin{equation}
\label{b8}
\eta_m ({\bf r}_i)=\lim_{t\to \infty}\int_{0}^{t}d\tau
\langle d(0)d(\tau)\rangle e^{i\omega_m\tau}
\end{equation}
is assumed to be the same for all detectors.

The ``self-efficiency'' $\eta_s$ of the Schr\"odinger detectors is
\begin{equation}
\label{b9}
\eta_s({\bf r}_i,{\bf r}_j)= \lim_{t\to \infty}\sum_{\nu} \Omega_\nu({\bf r}_i)
\Omega_\nu^{*}({\bf r}_j)
\int_{0}^{t} d\tau \langle {\bf \Psi}_{\nu}({\bf r}_j,0)
{\bf \Psi}_{\nu}^{\dagger}({\bf r}_i,\tau )\rangle
e^{i(\omega_L -\omega_s)\tau}
\end{equation}
with
\begin{equation}
\eta_s({\bf r}_i, {\bf r}_i) \equiv \eta_s({\bf r}_i),
\end{equation}
and their ``cross-efficiency'' is
\begin{eqnarray}
\eta_x ({\bf r}_i,{\bf r}_j)&=& 2\lim_{t \to \infty} \int_{0}^{t}d\tau
\int_{0}^{t}d\tau ' \int_{0}^{t}d\tau ''
e^{i(\omega_L -\omega_s )(\tau +\tau ')}
\sum_{\nu_1, \nu_2, {\bf q}_1 ,{\bf q}_2}
e^{-i[\omega_{{\bf q}_1\nu_1}\tau +\omega_{{\bf q}_2\nu_2}\tau']}
\nonumber \\
&\times& \left [e^{ -i(\omega_{\nu_1}-\omega_{\nu_2})(\tau - \tau '')}
|\Omega_{\nu_1}({\bf r}_i)|^2|\Omega_{\nu_2}({\bf r}_j)|^2
\phi_{{\bf q}_1}({\bf r}_i)\phi_{{\bf q}_1}^{\ast}({\bf r}_j )
\phi_{{\bf q}_2}({\bf r}_j )\phi_{{\bf q}_2}^{\ast}({\bf r}_i )\right .
\nonumber \\
&+& \left . e^{-i(\omega_{{\bf q}_1}-\omega_{{\bf q}_2})(\tau - \tau '')}
\Omega_{\nu_1}^{\ast}({\bf r}_i)\Omega_{\nu_1}({\bf r}_j)
\Omega_{\nu_2}^{\ast}({\bf r}_j)\Omega_{\nu_2}({\bf r}_i)|\phi_{{\bf q}_1}
({\bf r}_i)|^2|\phi_{{\bf q}_2 }({\bf r}_j)|^2 \right ] .
\label{b10}
\end{eqnarray}
In expression (\ref{b7}), the contribution of the Maxwell detectors is
the same as in conventional photodetection theory, as expected since we use
the same photoabsorption model. In contrast, the Schr\"odinger detectors
are responsible for three distinct terms, resulting from the fermionic
commutation relations of the electrons and ions. The first one, proportional
to $\eta_s({\bf r}_3)\eta_s({\bf r}_4)$, is of the same nature as the
expressions familiar from photodetection theory. It would take the same form in
case the ionizing fields at ${\bf r}_i$ and ${\bf r}_j$ lead to completely
distinguishable signals.

The second and third terms are of different nature, resulting from the quantum
statistics of the Schr\"odinger detectors. The first of these,
proportional to $\eta_s({\bf r}_i,{\bf r}_j)\eta_s({\bf r}_j,{\bf r}_i)$,
originates in the exchange of the ${\em composite}$ Schr\"odinger field
${\bf \Psi}_\nu({\bf r})$, which is bosonic. We note that the dependence
of this term on the product $\eta_s({\bf r}_i,{\bf r}_j)
\eta_s({\bf r}_j,{\bf r}_i)$  rather than on an expression proportional
to a fourth-order correlation function of ${\bf \Psi}_\nu({\bf r})$, is
a direct consequence of the assumption that this field is initially in a
vacuum. Finally, the third term, proportional to the cross-efficiency
$\eta_x$, involves the exchange of the electrons or of the ions {\em only}.
Physically, it accounts for the fact that the electrons don't know their origin.
This is the one term in Eq. (\ref{b10}) that depends explicitly on the
fact that the composite bosons are annihilated and a pair of fermions is
generated. The factor of 2 in $\eta_x$ results from the fact that either
the electrons or the ions can be exchanged. Hence, the first two terms can
be thought of as being bosonic in nature, while the latter term is fermionic.

We have previously mentioned that the $(N+M)$-th order transition probability
amplitude contains $(N+M)!$ terms. Therefore ${\cal P}(t)$ is made of $(4!)^2$
contributions, whose general form resembles that of expression (\ref{b7}).
Hence, it is readily possible in principle to make measurements sensitive
to both the bosonic and fermionic contributions to ${\cal P}(t)$. However,
if one is only interested in the electron counting rate
\begin{equation}
w( {\bf r}_1t_1,\ldots {\bf r}_4t_4)=
{\partial^4\over{\partial t_4\ldots\partial t_1}}
{\cal P}( {\bf r}_1t_1,\ldots {\bf r}_4t_4)
\label{w}
\end{equation}
then the specific ordering of $t_1, \ldots, t_4$ selects just one
of these contributions. In addition, since $w$ contains 4 partial derivatives,
the fermionic contribution to Eq. (\ref{b7}) gives no contribution to
the counting rate. Specifically, for the time ordering $t_1 <t_2<t_3<t_4$
one finds readily
\begin{eqnarray}
 &w&( {\bf r}_1 t_1,\ldots {\bf r}_4 t_4)=
\eta_m({\bf r}_1)\eta_m({\bf r}_2) \nonumber \\
&\times& \left \{\eta_s({\bf r}_3)\eta_s({\bf r}_4)
\langle {\bf E}^{(-)}({\bf r}_1,t_1)
{\bf E}^{(-)}({\bf r}_2,t_2)
 {\bf \Psi}^{\dagger}({\bf r}_3,t_3){\bf \Psi}^{\dagger}
({\bf r}_4,t_4)
{\bf \Psi}({\bf r}_4,t_4){\bf \Psi}({\bf r}_3,t_3)
{\bf E}^{(+)}({\bf r}_2,t_2)){\bf E}^{(+)}({\bf r}_1,t_1) \rangle \right .
\nonumber \\
&+& \left . \eta_s ({\bf r}_3,{\bf r}_4)\eta_s ({\bf r}_4 ,{\bf r}_3)
\langle {\bf E}^{(-)}({\bf r}_1,t_1)
{\bf E}^{(-)}({\bf r}_2,t_2)
{\bf \Psi}^{\dagger}({\bf r}_3,t_3){\bf \Psi}^{\dagger}({\bf r}_4,t_4)
{\bf \Psi}({\bf r}_4,t_3){\bf \Psi}({\bf r}_3,t_4)
{\bf E}^{(+)}({\bf r}_2,t_2){\bf E}^{(+)}({\bf r}_1,t_1)\rangle \right \}.
\label{b11}
\end{eqnarray}
The first term in this expression is analogous to the standard fourth-order
correlation function of optical coherence theory, with two of the electric
fields replaced by the Schr\"odinger field. The second term, which is as we
have seen an exchange term resulting from the undistingushability of the
detectors, can be eliminated if they are rendered distinguishable. This
could be realized experimentally in several ways, for instance by using
ionizaing lasers of different frequencies.\cite{thomas} In that case, the
counting rate reduces to a simple measure of the normally ordered
correlation function
\begin{eqnarray}
\label{b12}
G^{(N+M)}&(&x_1 ,...,x_{N+M})= \nonumber \\
&\langle &{\bf E}^{(-)}(x_1)...
{\bf E}^{(-)}(x_N)
{\bf \Psi}^{\dagger}(x_{N+1})...{\bf \Psi}^{\dagger}
(x_{N+M})
{\bf \Psi}(x_{N+M})...{\bf \Psi}(x_{N+1})
{\bf E}^{(+)}(x_N)...{\bf E}^{(+)}(x_1)\rangle ,
\end{eqnarray}
where $x_i=\{{\bf r}_i, t_i\}$.

We conclude this section by noting once more that different gating schemes
easily produce different cross-correlations functions between the optical
and the matter-wave fields. It is easily shown that the gating scheme always
selects the operators arrangement in such a way that time increases from the
outside to the inside in the resulting correlations functions. As a result,
of selecting the appropriate temporal ordering, it is possible to measure
correlations functions involving different combinations of light and matter
field operators. For example, the choice $t_3<t_1<t_4<t_2$ yields the
correlation function
\begin{equation}
\langle {\bf \Psi}^{\dagger}({\bf r}_3,t_3)  {\bf E}^{(-)}({\bf r}_1,t_1)
{\bf \Psi}^{\dagger}({\bf r}_4,t_4)  {\bf E}^{(-)}
({\bf r}_2,t_2)
{\bf E}^{(+)}({\bf r}_2,t_2)  {\bf \Psi} ({\bf r}_4,t_4)  {\bf E}^{(+)}
 ({\bf r}_1,t_1)  {\bf \Psi}({\bf r}_3,t_3)  \rangle .
\label{b13}
\end{equation}

\section{Example}

We now illustrate this theory by calculating selected third-order correlation
functions of a system consisting of linearly coupled optical and matter-wave
fields described by the effective Hamiltonian
\begin{equation}
\label{a1}
{\cal H}_0= \sum_{\alpha =1}^{R}\hbar \omega_{\alpha}a_{\alpha}^{\dagger}
a_{\alpha}+\sum_{i=1}^{S}\hbar \omega_{i}c_{i}^{\dagger}c_i+
\hbar\sum_{i \alpha}\left(  g_i^{(1)}a_{\alpha }^{\dagger}c_i+
 g_i^{(2)}a_{\alpha }c_i+ h.c.\right),
\end{equation}
where $R$ and $S$ are the number of modes in the optical and matter field,
respectively, with eigenfrequencies $\omega_{\alpha}$ and $\omega_{i}$.
In the following, we label the optical field with greek indices
and the matter waves with roman letters. The mode annihilation operators
\begin{equation}
\label{a2}
a_{\alpha}=\int d^3 r {\bf u}_{\alpha}^{*}({\bf r}) {\bf E}^{+}({\bf r})
\end{equation}
and
\begin{equation}
\label{a3}
c_{i}=\int d^3 r \phi_i^{*}({\bf r}) {\bf \Psi}({\bf r})
\end{equation}
are the bosonic optical and matter mode operators associated with the
mode functions ${\bf u}_{\alpha}({\bf r})$ and $phi_i({\bf r})$, respectively,
and $g_{i}^{(1)}$ and $g_{i}^{(2)}$ are coupling constants which allow
for a parametric amplification type of matter-light coupling as well.

The effective linear interaction of Eq. (\ref{a1}) while not
conserving the number of particles, is a good approximation to the
description of the coupling (for instance via a two-photon Raman transition)
between a macroscopically populated Bose-Einstein condensate ground state
and some weakly populated side-modes, see e.g. \cite{mike,zobay}. As long as
the condensate is not significantly depleted by that interaction, one can
invoke an ``undepleted'' mode approximation to replace the expectation value
of the associated field mode operator by a c-number order parameter,
leading to an interaction of the general form (\ref{a1}), with $g_{i}^{(1)}$
and $g_{i}^{(2)}$ proportional to the order parameter.

In addition to their Hamiltonian evolution, we assume that the matter and
light modes are also coupled to Markovian thermal reservoirs which result in
exponential losses at rates $\kappa_{\alpha}$ and $\kappa_{i}$.
The Langevin equations describing the evolution of this system are
\cite{Meystre}
\begin{equation}
\label{a4}
{d\over{dt}}{\bf x} (t)=-{\bf M}{\bf x} (t)+{\bf B}{\bf \xi}(t)
\end{equation}
where $\xi (t)$ are reservoir operators with $\langle \xi
(t)\rangle=0$ and
\begin{equation}
\langle \xi_i (t)\xi_j (t')\rangle= \delta_{ij}\delta(t-t').
\end{equation}
In the general expression (\ref{a4}) ${\bf x}(t)=
\{a_{\alpha}(t),c_{i}(t)\}^{T}$, $\alpha =\{1,...,R\}$;
$i=\{1,...,S\}$, and ${\bf M}$ is a $(R+S)\times (R+S)$ matrix whose
elements depend on the system parameters, and the difusion matrix
is ${\bf D}\equiv {\bf BB}^{\dagger}= diag (\bar{n}_{\alpha}
\kappa_{\alpha},\bar{n}_{i}\kappa_{i})$, where $\bar{n}_{\alpha}$ and
$\bar{n}_{i}$ are the thermal populations of the reservoir
modes.

We consider specifically the two-mode problem with drift matrix
\begin{equation}
{\bf M}=\pmatrix{\kappa_1/2& i g\cr
ig& \kappa_2/2},
\label{a7}
\end{equation}
and diffusion matrix
\begin{equation}
{\bf D}=\pmatrix{
\bar{n}_1\kappa_1& 0\cr
0&\bar{n}_1\kappa_2}
\label{a8}
\end{equation}
and evaluate the stationary value of the third-order correlation functions
\begin{equation}
\label{a5}
G^{(3)}_{x}= \langle {\bf \Psi}^{\dagger}({\bf r}_1,t_1)
{\bf \Psi}^{\dagger}({\bf r}_2,t_2){\bf E}^{(-)}({\bf r}_3,t_3)
{\bf E}^{(+)}({\bf r}_3,t_3){\bf \Psi}({\bf r}_2,t_2)
{\bf \Psi}({\bf r}_1,t_1) \rangle
\end{equation}
and
\begin{equation}
\label{a6}
G^{(3)}_{y}= \langle {\bf \Psi}^{\dagger}({\bf r}_1,t_1)
{\bf E}^{(-)}({\bf r}_3,t_2)
{\bf \Psi}^{\dagger}({\bf r}_2,t_3)
{\bf \Psi}({\bf r}_2,t_3){\bf E}^{(+)}({\bf r}_3,t_2)
{\bf \Psi}({\bf r}_1,t_1) \rangle  ,
\end{equation}
using the derivation outlined in Appendix A.

The normalized stationary cross-correlation function $g^{(3)}_{x}
(\tau_1,\tau_2)$ is shown in Fig. 1 as a function of the time differences
$\tau_1 =t_3 -t_1$ and $\tau_2=t_2 - t_1$. It corresponds to the time
gating $t_1<t_2<t_3$, so that $\tau_2 <\tau_1$. For the particular
parameters at hand, and for a fixed time difference $\tau_1$ we observe
a bunching-like behavior as a function of the time delay $\tau_2$ between
the Schr\"odinger detectors. This should be contrasted to Fig. 2, which
shows the normalized correlation function $g^{(3)}_y(\tau_1 ,\tau_2)$ for the
same parameters as in Fig. 1. In this case, $\tau_2$ is the time
difference between the photodetector and one of the Schr\"odinger
detectors, hence the behavior of the correlation function along $\tau_2$
exhibits a different behavior for fixed $\tau_1$. Note that the line
$\tau_1=\tau_2$ yields identical information in both cases, since it
corresponds to the identical physical situation where the optical detector
and one of the matter-wave detectors are simulataneously turned off.

\section{Summary and conclusions}

The dynamical interplay between Maxwell and Schr\"odinger waves, which is the
cornerstone of most novel applications of nonlinear and quantum atom optics,
requires one to be able to manipulate and characterize not just the
statistical properties of the optical and matter fields individually, but
also their cross-coherence properties.

In this paper, we have extended the theory of mutual coherence between
optical and matter waves to the study of higher-order correlation functions.
While in most studies of degenerate atomic quantum gases it is sufficient
to treat them as (composite) boson or fermion fields, this is no longer the
case if the detection mechanism destroys the particles, as is the case for
photoionization. The quantum statistics of the constituent particles play
an important role, as illustrated by Eqs. (\ref{b9}) and (\ref{b10}) which show a bosonic
term involving the simultaneous exchange of ions and electrons, and a
fermionic term involving the exchange of ions or electrons alone. While
it is possible to conceive specific counting schemes insensitive to these
terms, it would be extremely interesting to observe them.

The generalization of this analysis to fermionic fields is also
of much interest, since in this case, the constituents involved are
a fermionic field and a bosonic field instead of two fermionic fields as
was the case in this paper. The exchange contributions to the ionization
probability will be quite different in that case, since electron exchange,
as well as the simultaneous exchange of ions and electrons, willl now lead
to fermionic contributions, while the exchange of ions produces a bosonic term.

\acknowledgments{This work is supported in part by Office of Naval Research
Contract No. 14-91-J1205, National Science Foundation Grant PHY-9801099,
the Army Research Office and the Joint Services Optics Program.
JZ thanks COLCIENCIAS, Universidad del Atl\'antico and Universidad del Norte,
and GAP thanks FAPESP (Funda\c c\~ao de Amparo a Pesquisa do Estado de
S\~ao Paulo), for financial support. Discussions with E. V. Goldstein,
M. G. Moore, J. Heurich and E. M. Wright are greatfully acknowledged.}

\appendix
\section{correlation functions matrix elements}

From the expansions (\ref{a2}) and (\ref{a3}) of the Schr\"odinger field
operators, the correlation function $G^{(3)}_x$ is
\begin{eqnarray}
\label{ap1}
G^{(3)}_x &=&\sum_{ij\alpha\beta kl}
\phi^{*}_i({\bf r}_1)\phi^{*}_j({\bf r}_2)
{\bf u}^{*}_{\alpha}({\bf r}_3){\bf u}_{\beta}({\bf r}_3)
\phi_k({\bf r}_2)\phi_l({\bf r}_1)\nonumber \\
&\{ &\left(U^{-1}{\cal G}(t_3,t_1)(U^{-1})^{\dagger}\right)_{\beta i}
\left(U^{-1}{\cal G}(t_2,t_2)(U^{-1})^{\dagger}\right)_{k j}
\left(U^{-1}{\cal G}(t_1,t_3)(U^{-1})^{\dagger}\right)_{l \alpha }\nonumber \\
&+ &\left(U^{-1}{\cal G}(t_3,t_1)(U^{-1})^{\dagger}\right)_{\beta i}
\left(U^{-1}{\cal G}(t_1,t_2)(U^{-1})^{\dagger}\right)_{l j}
\left(U^{-1}{\cal G}(t_2,t_3)(U^{-1})^{\dagger}\right)_{k \alpha }\nonumber \\
&+ &\left(U^{-1}{\cal G}(t_2,t_1)(U^{-1})^{\dagger}\right)_{k i}
\left(U^{-1}{\cal G}(t_3,t_2)(U^{-1})^{\dagger}\right)_{\beta j}
\left(U^{-1}{\cal G}(t_1,t_3)(U^{-1})^{\dagger}\right)_{l \alpha }\nonumber \\
&+&\left(U^{-1}{\cal G}(t_2,t_1)(U^{-1})^{\dagger}\right)_{k i}
\left(U^{-1}{\cal G}(t_1,t_2)(U^{-1})^{\dagger}\right)_{l j}
\left(U^{-1}{\cal G}(t_3,t_3)(U^{-1})^{\dagger}\right)_{\beta \alpha }
\nonumber \\
&+&\left(U^{-1}{\cal G}(t_1,t_1)(U^{-1})^{\dagger}\right)_{l i}
\left(U^{-1}{\cal G}(t_3,t_2)(U^{-1})^{\dagger}\right)_{\beta j}
\left(U^{-1}{\cal G}(t_2,t_3)(U^{-1})^{\dagger}\right)_{k \alpha }\nonumber \\
&+&\left(U^{-1}{\cal G}(t_1,t_1)(U^{-1})^{\dagger}\right)_{l i}
\left(U^{-1}{\cal G}(t_2,t_2)(U^{-1})^{\dagger}\right)_{k j}
\left(U^{-1}{\cal G}(t_3,t_3)(U^{-1})^{\dagger}\right)_{\beta \alpha }\}
\end{eqnarray}
where $U$ is the matrix of eigenvectors of the system
matrix ${\bf M}$, so that
\begin{equation}
\label{ap2}
U^{-1}{\bf M}U= diag(\lambda_k ),\,\, k=\{1,...,R+S \},
\end{equation}
where $\{ \lambda_k \}$ are the eigenvalues of ${\bf M}$,
and
\begin{eqnarray}
\label{ap3}
{\cal G}_{ij}(t,s)&=& (U{\bf D}U^{\dagger})_{ij}
{e^{-\lambda_i (t-s)}\over{\lambda_i+\lambda_j^{*}}}\,\,\,\,\, t\ge s,
\nonumber \\
{\cal G}_{ij}(t,s)&=& (U{\bf D}U^{\dagger})_{ij}
{e^{-\lambda_j^{*} (s-t)}\over{\lambda_i+\lambda_j^{*}}}\,\,\,\,\, t< s.
\end{eqnarray}
The correlation $G^{(3)}_y$ has a similar expression.

The six terms in the Eq. (\ref{ap1}) follow directly from
the steady state solution of Eq. (\ref{a4}) given by
\begin{equation}
\label{ap4}
{\bf x}(t)=\int_{-\infty}^{t} e^{-{\bf M}(t-t')}{\bf B\xi}(t')dt'.
\end{equation}

--------------------------------------------------------------------

\newpage

\begin{figure}
\includegraphics[width=1.0\columnwidth]{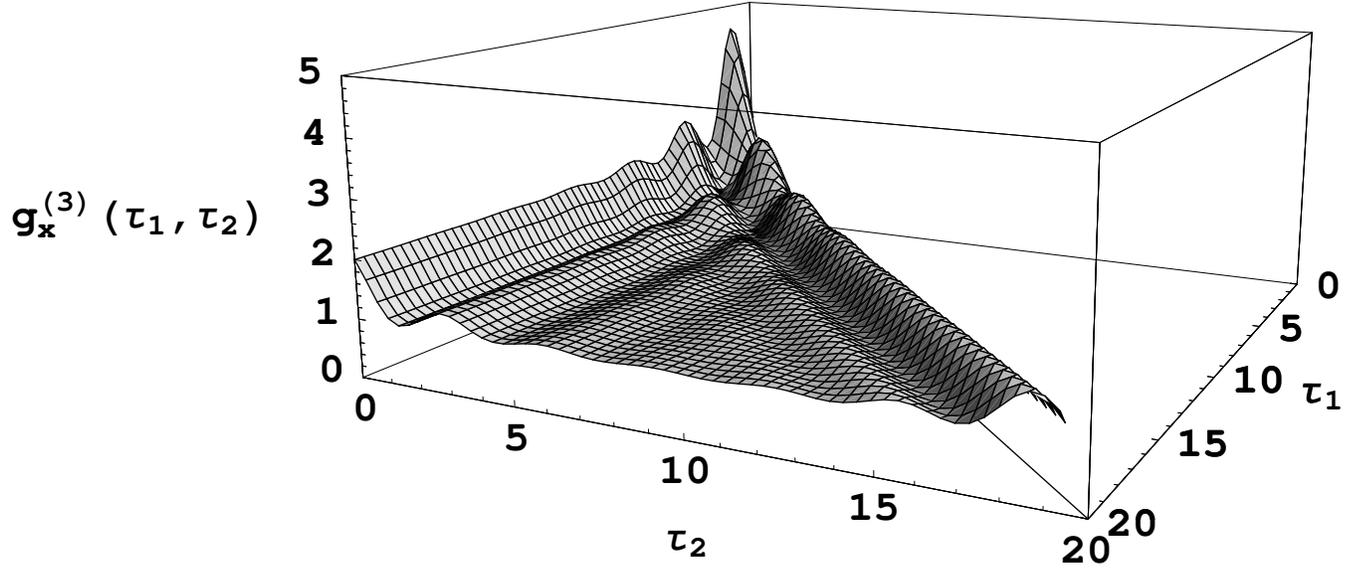}
\caption{Normalized correlation function $g^{(3)}_x (\tau_1,\tau_2)$
as a function of $\tau_1 =t_3 -t_1 $ and $\tau_2 =t_2 -t_1$ for $\kappa_1 = 0.15$,
$\kappa_2 = 0.25$, $\bar{n}_1=0.01$ and $\bar{n}_2 =0.1$, in units of the coupling 
constant $g$.}
\end{figure}

\newpage

\begin{figure}
\includegraphics[width=1.0\columnwidth]{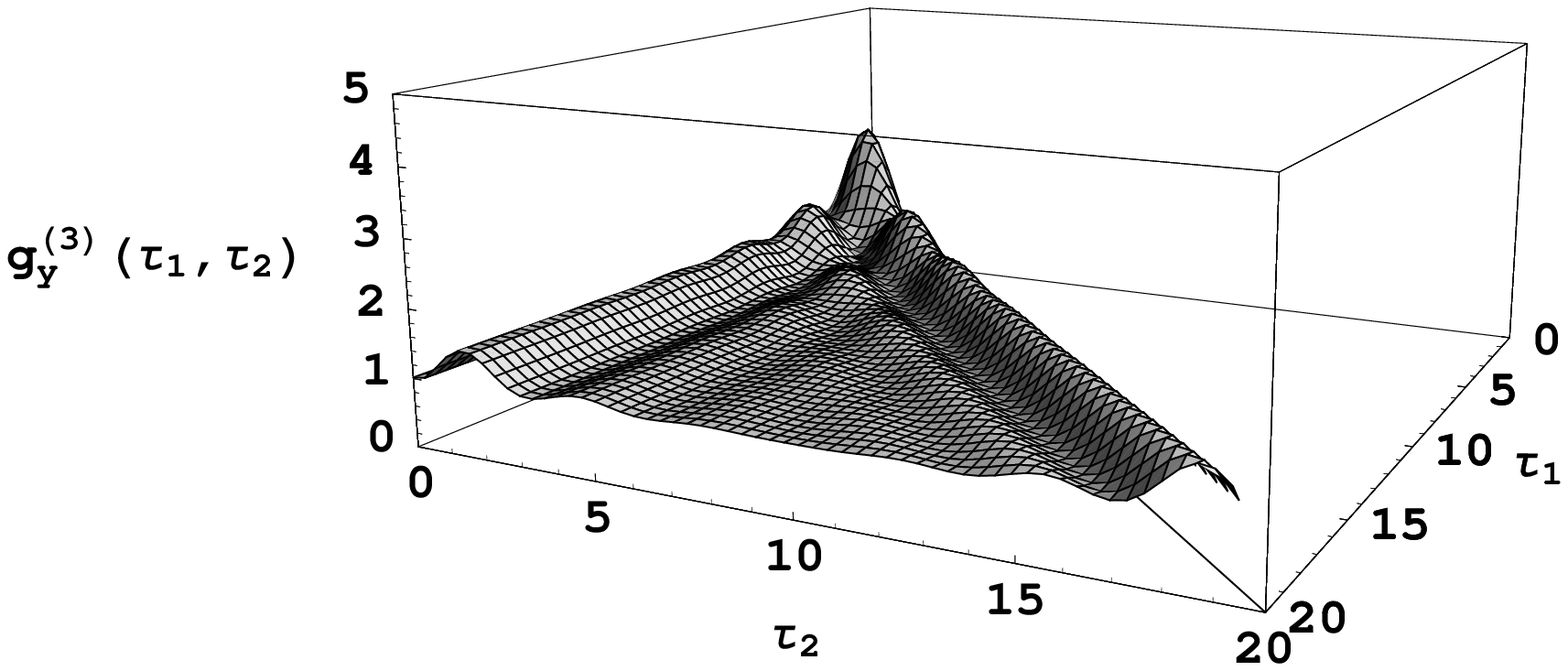}
\caption{Same as Fig. 1, for $g^{(3)}_{y}(\tau_1,\tau_2)$.}
\end{figure}


\begin{references}
\bibitem{r1} M. H. Anderson, J. R. Ensher, M. R. Mathews, C. E. Wieman, and E.
A. Cornell, Science {\bf 269}, 198 {1995}.
\bibitem{r2} K. B. Davis, M. -O. Mewes, M. R. Andrews, N. J. van Drutem,
D. S. Durfee, D. M. Kurn, and W. Ketterle, Phys. Rev. Lett. {\bf 75}, 3969
(1995).
\bibitem{r3} J. R. Ensher, D. S. Jin, M. R. Mathews, C. E. Wieman, and
E. A. Cornell, Phyis. Rev. Lett. {\bf 77}, 4984 (1996).
\bibitem{r4} M. -O. Mewes, M. R. Andrews, N. J. van Drutem,
D. S. Durfee, D. M. Kurn, and W. Ketterle, Phys. Rev. Lett. {\bf 77}, 416
(1996).
\bibitem{r5} C. C. Bradley, C. A. Sackett, and R. G. Hulet, Phys. Rev. Lett.
{\bf 78}, 985 (1997).
\bibitem{r6} L. Deng {\it et al.}, Nature (London) {\bf 398}, 218 (1999).
\bibitem{ert99} S. Burger {\it et al}, cond-mat/9910487, to be published in
Phys. Rev. Lett.
\bibitem{phi99} W. D. Phillips, S. L. Rolston et al, unpublished (1999).
\bibitem{mew97} M. O. Mewes et al, Phys. Rev. Lett. {\bf 78}, 582 (1997).
\bibitem{and98} B. P. Anderson and M. A. Kasevich, Science {\bf 282}, 1686 (1998).
\bibitem{r7} E. W. Hagley {\it et al.}, Science {\bf 283},1706 (1999).
\bibitem{r8} I. Bloch, T. W. H\"ansch, T. Esslinger, Phys. Rev. Lett. {\bf 82},
3008 (1999).
\bibitem{r9} S. Inouye {\it et al.} Science {\bf 285}, 571 (1999).
\bibitem{r10} G. A. Prataviera, E. V. Goldstein, and P. Meystre,
Phys. Rev. A {\bf 60}, 4846 (1999).
\bibitem{r11} D. F. Walls and G. J. Milburn, {\it Quantum Optics}
(Springer-Verlag, Berlin, 1994).
\bibitem{BouKas97} P. Bouyer and M. A. Kasevich, Phys. Rev. A {\bf 56},
R1083 (1997).
\bibitem{ket99z} S. Inouye et al, preprint (1999).
\bibitem{elena} E. V. Goldstein and P. Meystre, Phys. Rev. Lett. {\bf 80},
5036 (1998).
\bibitem{glauber}  R. J. Glauber, in {\it Quantum Optics and
Electronics}, edited C. de Witt, A. Blandin, and C.
Cohen-Tannoudji (Gordon and Breach, New York, 1965).
\bibitem{thomas} J. E. Thomas and L. J. Wang, Phys. Rev. A {\bf 49}, 558 (1994).
\bibitem{cohen}  C. Cohen-Tannoudji, J. Dupont-Roc, and G. Grynberg, {\it
Atom-Photon Interactions}, Wiley-Interscience (New York, 1992).
\bibitem{mike} M. G. Moore and P. Meystre, Phys. Rev. A {\bf 59},
R1754 (1999).
\bibitem{zobay} M. G. Moore, O. Zobay, and P. Meystre, Phys. Rev. A {\bf 60},
 1491 (1999).
\bibitem{Meystre} P. Meystre and M. Sargent III; {\it Elements
of Quantum Optics}, 3rd Edition (Springer-Verlag, Heidelberg, 1998).
\end{references}
\end{document}